\documentstyle[11pt,dunk2001_asp,twoside,psfig]{article}
\begin{document}

\def\emphasize#1{{\sl#1\/}}
\def\arg#1{{\it#1\/}}
\let\prog=\arg

\def\edcomment#1{\iffalse\marginpar{\raggedright\sl#1\/}\else\relax\fi}
\marginparwidth 1.25in
\marginparsep .125in
\marginparpush .25in
\reversemarginpar

\def\kms {\ km s$^{-1}$}
\def\msol{\ifmmode {\>M_\odot}\else {$M_\odot$}\fi}
\def\cmsq{\ifmmode {\>{\rm\ cm}^2}\else {cm$^2$}\fi}
\def\psqcm{\ifmmode {\>{\rm cm}^{-2}}\else {cm$^{-2}$}\fi}
\def\psqpc{\ifmmode {\>{\rm pc}^{-2}}\else {pc$^{-2}$}\fi}
\def\pcsq{\ifmmode {\>{\rm\ pc}^2}\else {pc$^2$}\fi}
\def\Tkev{\ifmmode{T_{\rm kev}}\else {$T_{\rm keV}$}\fi}
\def\hubunits{\ifmmode {\>{\rm km\ s^{-1}\ Mpc^{-1}}}\else {km
s$^{-1}$ Mpc$^{-1}$}\fi}
\def\gta{\;\lower 0.5ex\hbox{$\buildrel > \over \sim\ $}}
\def\lta{\;\lower 0.5ex\hbox{$\buildrel < \over \sim\ $}}

\def\phiIV{\ifmmode{\varphi_4}\else {$\varphi_4$}\fi}
\def\phiI{\ifmmode{\varphi_i}\else {$\varphi_i$}\fi}

\def\be{\begin{equation}}
\def\ee{\end{equation}}
\def\bea{\begin{eqnarray}}
\def\eea{\end{eqnarray}}
\def\beas{\begin{eqnarray*}}
\def\eeas{\end{eqnarray*}}
\def\gtrapprox{\;\lower 0.5ex\hbox{$\buildrel >\over \sim\ $}}
\def\lessapprox{\;\lower 0.5ex\hbox{$\buildrel < \over \sim\ $}}
\def\deg   {$^\circ$}
\def\Ftwo  {$F_{-21}$}
\def\Pcos  {$\Phi^0$}
\def\Jtwo  {$J_{-21}$}
\def\Fcos  {$F_{-21}^0$}
\def\Jcos  {$J_{-21}^0$}
\def\Em    {${\cal E}_m$}
\def\ALL   {A_{\scriptscriptstyle LL}}
\def\JLL   {J_{\scriptscriptstyle LL}}
\def\nuLL  {\nu_{\scriptscriptstyle LL}}
\def\sigLL {\sigma_{\scriptscriptstyle LL}}
\def\tauLL {\ifmmode{\tau_{\scriptscriptstyle LL}}\else 
           {$\tau_{\scriptscriptstyle LL}$}\fi}
\def\nuOB  {\nu_{\scriptscriptstyle {\rm OB}}}
\def\aB    {\alpha_{\scriptscriptstyle B}}
\def\nH    {n_{\scriptscriptstyle H}}
\def\Em{\ifmmode{{\rm E}_m}\else {{\rm E}$_m$}\fi}
\def\NH{\ifmmode{{\rm N}_{\scriptscriptstyle\rm H}}\else {{\rm N}$_{\scriptscriptstyle\rm H}$}\fi}

\def\Ha    {H$\alpha$}
\def\Hb    {H$\beta$}
\def\HI    {H${\scriptstyle\rm I}$}
\def\HII    {H${\scriptstyle\rm II}$}
\def\eg    {{\it e.g.}}
\def\ie    {{\it i.e.}}
\def\cf    {{\it cf. }}
\def\qv    {{\it q.v. }}
\def\etal  {\ et al.}
\def\kms{\ifmmode {\>{\rm\ km\ s}^{-1}}\else {\ km s$^{-1}$}\fi}

\def\Em{\ifmmode{{\cal E}_m}\else {{\cal E}$_m$}\fi}
\def\Dm{\ifmmode{{\cal D}_m}\else {{\cal D}$_m$}\fi}
\def\fesc{\ifmmode{\hat{f}_{\rm esc}}\else {$\hat{f}_{\rm esc}$}\fi}
\def\fescs{\ifmmode{f_{\rm esc}}\else {$f_{\rm esc}$}\fi}
\def\rsolar{\ifmmode{r_\odot}\else {$r_\odot$}\fi}
\def\emunit{\ifmmode{{\rm cm}^{-6}{\rm\ pc}}\else {
cm$^{-6}$ pc}\fi}
\def\intensity{\ifmmode{{\rm erg\ cm}^{-2}{\rm\ s}^{-1}
      {\rm\ Hz}^{-1}{\rm\ sr}^{-1}}
      \else {erg cm$^{-2}$ s$^{-1}$ Hz$^{-1}$ sr$^{-1}$}\fi}
\def\flux{\ifmmode{{\rm erg\ cm}^{-2}{\rm\ s}^{-1}}\else {erg
cm$^{-2}$ s$^{-1}$}\fi}
\def\fluxdensity{\ifmmode{{\rm erg\ cm^{-2}\ s^{-1}\ Hz^{-1}}}\else {erg
cm$^{-2}$ s$^{-1}$ Hz$^{-1}$}\fi}
\def\phoflux{\ifmmode{{\rm phot\ cm}^{-2}{\rm\ s}^{-1}}\else {phot
cm$^{-2}$ s$^{-1}$}\fi}
\def\phorate{\ifmmode{{\rm phot\ s}^{-1}}\else {phot s$^{-1}$}\fi}

\def\apj{{\it Ap.J.~}}
\def\apjs{{\it Ap.J.Suppl.~}}
\def\apss{{\it Astrophys.Sp.Science~}}
\def\aj{{\it Astron.J.~}}
\def\aph{{\it astro-ph}}
\def\mn{{\it MNRAS~}}
\def\araa{{\it Annu.Rev.Astron.Astrophys.~}}
\def\pasp{{\it PASP.~}}
\def\aaa{{\it Astron.Astrophys.~}}
\def\aaas{{\it Astron.Astrophys.Suppl.~}}
\def\astroph{{\it astro-ph}}
\def\rmp{{\it Rev.Mod.Phys.~}}

\title{\bf The Gaiasphere and the limits of knowledge}

\author{Joss Bland-Hawthorn} 
\affil{Anglo-Australian Observatory, 167 Vimiera Road, Eastwood, NSW 2122, Australia; jbh@aao.gov.au}

\begin{abstract}
At the heart of a successful theory of galaxy formation must be
a detailed physical understanding of the dissipational processes which
form spiral galaxies.  To what extent can we unravel the events that
produced the Galaxy as we see it today?  Could some of the residual
inhomogeneities from prehistory have escaped the dissipative process
at an early stage?  To make a comprehensive inventory of surviving
inhomogeneities would require a vast catalog of stellar properties that is
presently out of reach.  The Gaia space astrometry mission, set to launch
at the end of the decade, will acquire detailed phase space coordinates
for about one billion stars, within a sphere of diameter 20 kpc --
the Gaiasphere.  Here we look forward to a time when all stars within
the Gaiasphere have complete chemical abundance measurements (including
$\alpha$, $s$ and $r$ process elements).  Even with such a vast increase
in information, there may exist fundamental $-$ but unproven $-$ limits
to unravelling the observed complexity.
\end{abstract}

\vspace{.5in}

\section{Introduction}
Eddington once remarked that `the contemplation in natural science of
a wider domain than the actual leads to a far better understanding of
the actual.' Our intuition is that any dynamical phase early on in the
history of the Galaxy which is dominated by relaxation or dissipation
probably removes more information than it retains.  This remains largely
true for controlled experiments within terrestrial laboratories for the
reason that it is exceedingly difficult to track fluid particles,
for example, in order to unravel the process under study.  However, at
least in principle, the situation may be more tractable in astrophysics
in that individual stars carry information about the birth site at the
time of their formation.  Nature offers us a vast untapped reservoir of
information in the detailed chemical abundances stored within stellar
atmospheres.

The fossil evidence of galaxy formation is not restricted to chemical
abundances alone. In our review, ``The New Galaxy -- Signatures
of its Formation,'' we stress the importance of new observations across a
wide range of parameter space (Freeman \& Bland-Hawthorn 2002).  We are
coming into a new era of galactic investigation, in which one can study
the fossil remnants of the early days of the Galaxy in a broader and more
focussed way, not only in the halo but throughout the major luminous
components of the Galaxy.  The goal of these studies is to reconstruct
as much as possible of the early galactic history.

What do we mean by the reconstruction of early galactic history? We
seek a detailed physical understanding of the sequence of events which
led to the Milky Way. Ideally, we would want to tag (i.e. associate)
components of the Galaxy to elements of the protocloud -- the baryon
reservoir which fueled the stars in the Galaxy.  

\begin{figure}
\psfig{file=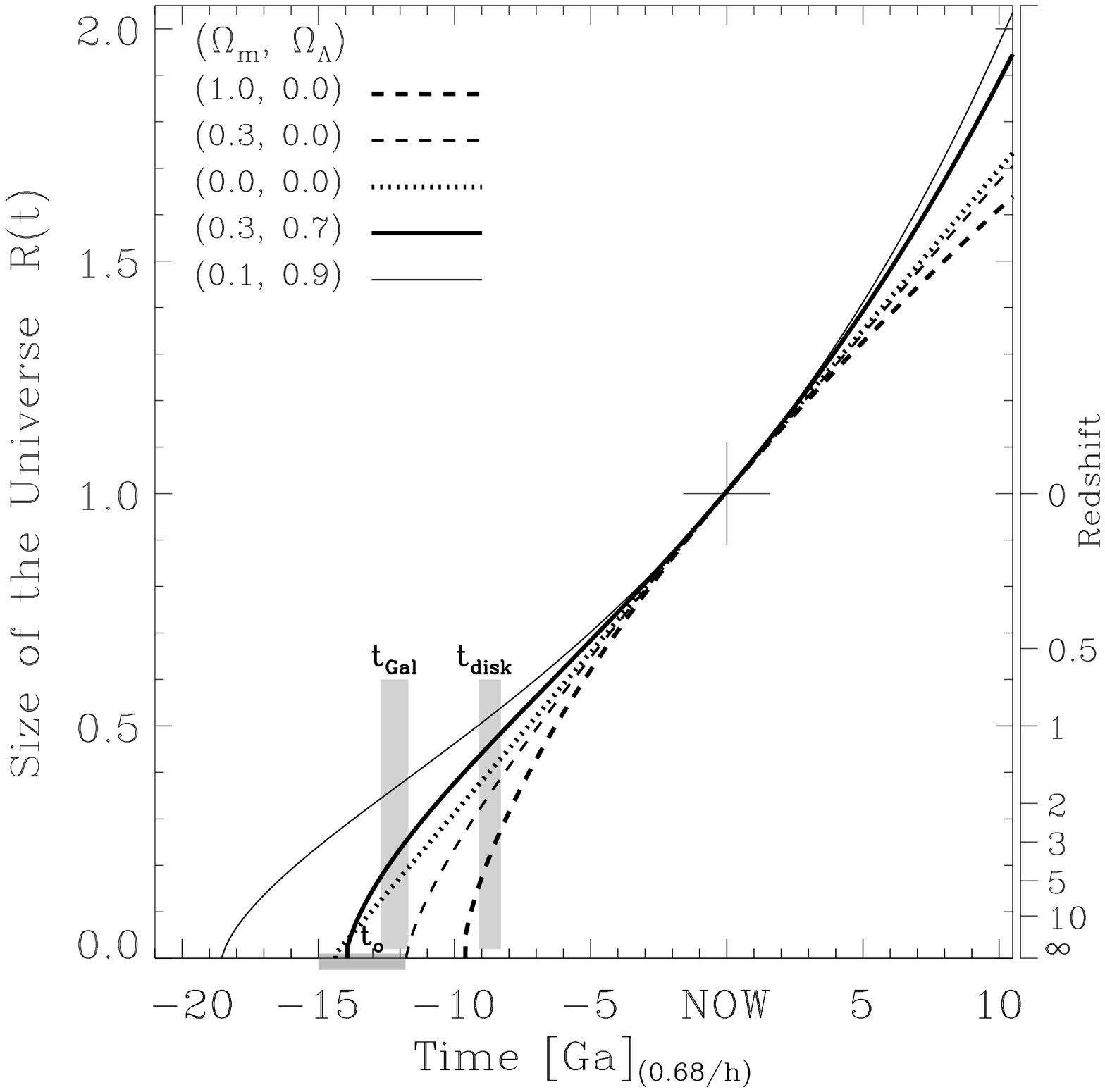,width=4.8in}
\caption{
Look-back time as a function of redshift and the size of the Universe
(Lineweaver 1999) for five different world models. The approximate ages
of the Galactic halo and disk are indicated by hatched regions. 
\label{fig1}
}
\end{figure}

Our approach is anchored to observations of the Galaxy,
interpreted within the broad scope of the CDM hierarchy. Many of the
observables in the Galaxy relate to events which occurred long ago, at
high redshift. Fig.~\ref{fig1} shows the relationship between look-back
time and redshift in the context of the $\Lambda$CDM model: the
redshift range (z $\lta$ 6) of discrete sources in contemporary
observational cosmology matches closely the known ages of the oldest
components in the Galaxy.  The Galaxy (near-field cosmology) provides a
link to the distant universe (far-field cosmology).

Here we address the prospect of chemically `tagging' stars to common
sites of formation, in order to reconstruct star clusters which have
long since dispersed.  High resolution spectroscopy at high
signal-to-noise ratio of many stellar types reveals an extraordinarily
complex pattern of spectral lines.  The spectral lines carry key
information on element abundances that make up the stellar atmosphere.
Some of these elements cannot arise through normal stellar evolution,
and therefore must reflect conditions in the progenitor cloud at the
time of its formation. A cornerstone of near-field cosmology must be to
explore how much of the Galaxy's past can be reconstructed from the
chemical signatures in stellar atmospheres.

\section{Chemical signatures}

A major goal of near-field cosmology is to tag individual stars with
elements of the protocloud.  For many halo stars, and some outer bulge
stars, this may be possible with phase space information provided by
Gaia. But for much of the bulge and the disk, secular processes cause
the populations to become relaxed (\ie\ the integrals of motion are
partially randomized).  In order to have any chance of unravelling disk
formation, we must explore chemical signatures in the stellar
spectrum.  Ideally, we would like to tag a large sample of
representative stars with a precise {\it time} and a precise {\it site}
of formation.

\begin{figure}
\psfig{file=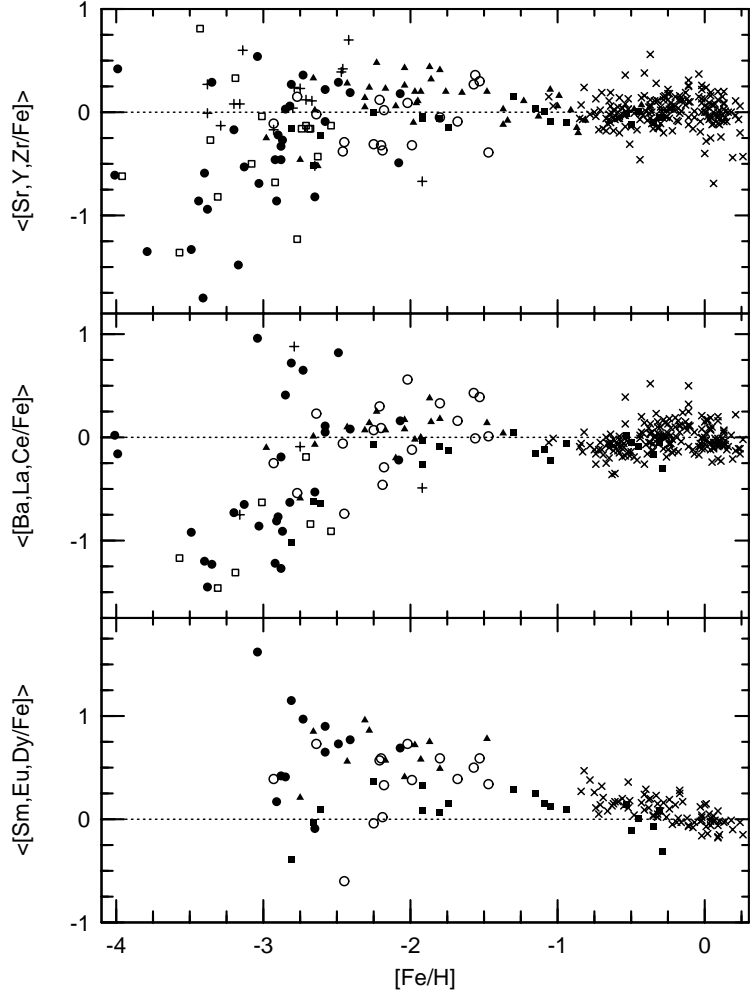,width=4.8in}
\caption{
Mean relative abundance ratios of light s-process elements (top
panel), heavy s-process elements (middle panel), and and r-process
elements (bottom panel) as functions of [Fe/H]. In each panel, 
the dotted horizontal lines represent the solar abundance ratios of 
these elements. The references for the data points are given in
Wallerstein\etal\ (1997). [We thank C. Sneden for preparing this
figure.]
\label{fig2}
}
\end{figure}

Over the last four decades, evidence has gradually accumulated
(Fig.~\ref{fig2}) for a large dispersion in metal abundances [X$_i$/Fe]
(particularly n-capture elements) in low metallicity stars relative to
solar abundances (Wallerstein\etal\ 1963; Pagel 1965; Spite \& Spite
1978; Truran 1981; Luck \& Bond 1985; Clayton 1988; Gilroy\etal\ 1988;
McWilliam\etal\ 1995; Norris \etal\ 1996; Burris\etal\ 2000).  Elements
like Sr, Ba and Eu show a 300-fold dispersion, although [$\alpha$/Fe]
dispersions are typically an order of magnitude smaller.

In their celebrated paper, Burbidge\etal\ (1957 $-$ B$^2$FH)
demonstrated the likely sites for the synthesis of slow (s) and rapid
(r) n-capture elements. The s-process elements (\eg\ Sr, Zr, Ba, Ce,
La, Pb) are thought to arise from the He-burning phase of
intermediate to low mass (AGB) stars (M $<$ 10M$_\odot$), although
at the lowest metallicities, trace amounts are likely to arise from
high mass stars (Burris\etal\ 2000; Rauscher\etal\ 2001).

In contrast to the s-process elements, the r-process elements (\eg\ Sm,
Eu, Gd, Tb, Dy, Ho) cannot be formed during quiescent stellar
evolution.  While some doubts remain, the most likely site for the
r-process appears to be SN~II, as originally suggested by B$^2$FH (see
also Wallerstein\etal\ 1997). Therefore, r-process elements measured
from stellar atmospheres reflect conditions in the progenitor cloud. In
support of Gilroy\etal\ (1988), McWilliam\etal\ (1995) state that
`the very large scatter means that n-capture element abundances in
ultra-metal poor stars are products of one or very few prior
nucleosynthesis events that occurred in the very early, poorly mixed
galactic halo', a theme that has been developed by many authors
(\eg\ Audouze \& Silk 1995; Shigeyama \& Tsujimoto 1998; Argast\etal\
2000; Tsujimoto \etal\ 2000).

Supernova models produce different yields as a function
of progenitor mass, progenitor metallicity, mass cut (what gets
ejected compared to what falls back towards the compact central
object), and detonation details. The $\alpha$ elements are mainly 
produced in the hydrostatic burning phase within the pre-supernova star.  
Thus $\alpha$ yields are not dependent on the mass cut or
details of the fallback/explosion mechanism which leads to a much
smaller dispersion at low metallicity.

There is no known age-metallicity relation that operates over a useful
dynamic range in age and/or metallicity. (This effect is only seen in a
small subset of hot metal-rich stars).  Such a relation would require the
metals to be well mixed over large volumes of the ISM.  For the
forseeable future, it seems that only a small fraction of stars can be
dated directly.  Arguably, the most promising stellar chronometers
make use of radioactive dating or asteroseismology.
Nucleo-cosmochronology is not yet a precise science but has potential
for future development (Sneden\etal\ 2001$a$; Goriely \& Arnould
2001).

\section{Reconstructing ancient star groups}

We now conjecture that the heavy element metallicity dispersion may
provide a way forward for tagging groups of stars to common sites of
formation.  With sufficiently detailed spectral line information, it is
feasible that the `chemical tagging' will allow temporal sequencing of
a large fraction of stars in a manner analogous to building a family
tree through DNA sequencing.

Most stars are born within rich clusters of many hundreds to many
thousands of stars (Clarke\etal\ 2000; Carpenter 2000).  McKee \& Tan
(2002) propose that high-mass stars form in the cores of strongly
self-gravitating and turbulent gas clouds. The low mass stars form
within the cloud outside the core, presumably at about the same time or
shortly after the high mass stars have formed. The precise sequence of
events which give rise to a high mass star is a topic of great interest
and heated debate in contemporary astrophysics
(\eg\ Stahler\etal\ 2000).

A necessary condition for `chemical tagging' is that the progenitor
cloud is uniformly mixed in key chemical elements before the first stars
are formed.  Another possibility is that a few high mass stars form
shortly after the cloud assembles, and enrich the cloud fairly uniformly.
Both scenarios would help to ensure that long-lived stars have identical
abundances in certain key elements before the onset of low-mass star
formation.

For either statement to be true, an important requirement is that {\it open
clusters of any age have essentially zero dispersion in some key metals
with respect to Fe}.  There has been very little work on heavy element
abundances in open clusters. The target clusters must have reliable
astrometry so as to minimize `pollution' from stars not associated with
the cluster (Quillen 2002).

If our requirement is found not to be true, then either the progenitor
clouds are not well mixed or high mass stars are formed after most
low mass stars. A more fundamental consequence is that {\it a direct
unravelling of the disk into its constituent star groups would be
impossible, in other words, the epoch of dissipation cannot be unravelled
after all.}

Consider the (extraordinary) possibility that we {\it could} put
many coeval star groups back together over the entire age of the Galaxy.
This would provide an accurate age for the star groups either through
the color-magnitude diagram, or through association with those stars
within each group that have [n-capture/Fe] $\gg$ 0, and can therefore
be radioactively dated.  This would provide key information on the
chemical evolution history for each of the main components of the
Galaxy.

But what about the formation site? The kinematic signatures will
identify which component of the Galaxy the reconstructed star group
belongs, but not specifically where in the Galactic component
(\eg\ radius) the star group came into existence. For stars in the thin
disk and bulge, the stellar kinematics will have been much affected by
the bar and spiral waves; it will no longer be possible to estimate
their birthplace from their kinematics.  Our expectation is that the
derived family tree will severely restrict the possible scenarios
involved in the dissipation process. In this respect, a sufficiently
detailed model may be able to locate each star group within the
simulated time sequence.

Freeman \& Bland-Hawthorn (2002) argue that, in addition to open
clusters, the thick disk is an extremely important fossil of the
processes behind disk formation. The thick disk is thought to be 
a snap-frozen relic of the early disk, heated vertically by the
the infall of an intermediate mass satellite. Chemical tagging
of stars that make up the thick disk would provide clues on the 
formation of the first star clusters in the early disk.

\section{Chemical abundance space}

An intriguing prospect is that reconstructed star clusters can be
placed into an evolutionary sequence, \ie\ a family tree, based on
their chemical signatures. Let us suppose that a star cluster has
accurate chemical abundances determined for a large number $n$ of
elements (including isotopes). This gives it a unique location in an
$n$-dimensional space compared to $m$ other star clusters within that
space. We write the chemical abundance space as
${\cal C}$(Fe/H, X$_1$/Fe, X$_2$/Fe, ...) where X$_1$, X$_2$ ... 
are the independent chemical elements that define the space (\ie\
elements whose abundances are not rigidly coupled to other elements).

\begin{figure}
\psfig{file=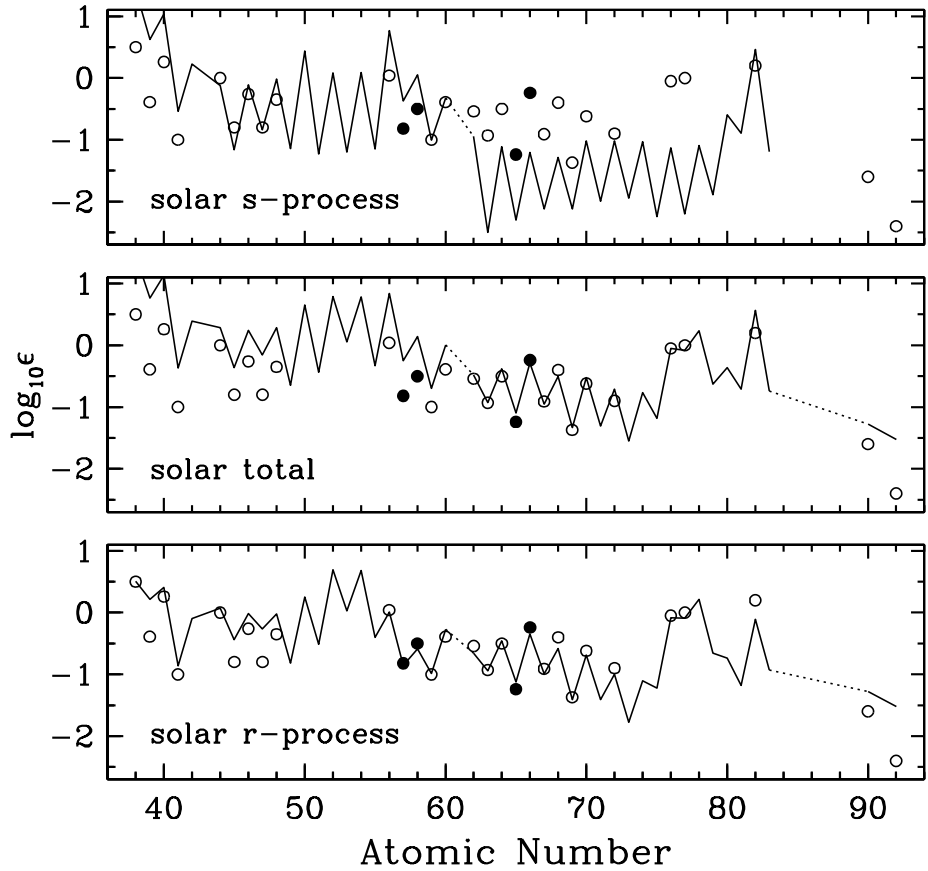,width=4.8in}
\caption{
CS 22892-052 n-capture abundances (points) taken from Sneden\etal\ 
(2000) and {\it scaled} solar system abundances  (solid and dashed lines) 
taken from Burris\etal\ (2000). Many of the heavy elements conform to
the solar system r-process abundance pattern, although some elements show 
the hallmark of the s-process. This figure was originally presented
in Sneden\etal\ (2001$a$).
\label{fig3}
}
\end{figure}

The size of $n$ is unlikely to exceed about 50 for the foreseeable
future. Hill\etal\ (2002) present exquisite data for the metal-poor
star CS~31082-001, where abundance estimates are obtained for a total
of 44 elements, almost half the entire periodic table (see also
Cayrel\etal\ 2001).  In Fig.\ref{fig3}, we show what is now possible
for another metal-poor star, CS~22892-052 (Sneden\etal\ 2001$a$).
The $\alpha$ elements and r-process elements, and
maybe a few canonical s-process elements at low [Fe/H], provide
information on the cloud abundances prior to star formation, although
combinations of these are likely to be coupled (Heger \& Woosley 2001;
Sneden\etal\ 2001$a$).  There are 24 r-process elements that have been
clearly identified in stellar spectra (Wallerstein\etal\ 1997).

The size of $m$ is likely to be exceedingly large for the thin disk
where most of the baryons reside. For a rough estimate, we take the age
of the disk to be 10~Ga. If there is a unique SN~II enrichment event
every 100 years, we expect of order 10$^8$ formation sites.  Typically,
a SN~II event sweeps up a constant mass of $5\times 10^4$M$_\odot$
(Ryan\etal\ 1996; Shigeyama \& Tsujimoto 1998). Simple chemical
evolution models indicate that this must be of the right order to
explain the metallicity dispersion at low [Fe/H] (Argast\etal\ 2000).
Roughly speaking, there have been 10$^3$ generations of clouds since
the disk formed, with about 10$^5$ clouds in each star-forming
generation, such that cloud formation and dispersal cycles on a 10$^7$
yr timescale (Elmegreen\etal\ 2000).

Whereas the total number of star clusters over the lifetime of the thin
disk is very large, the size of $m$ for the stellar halo
(Harding\etal\ 2001), and maybe the thick disk (Kroupa 2002), is likely
to be significantly smaller.  Our primary interest is the oldest star
clusters.  Reconstructing star clusters within the thick disk is a
particularly interesting prospect since the disk is likely to have
formed within 1$-$1.5~Ga of the main epoch of baryon dissipation
(Prochaska\etal\ 2000).

The task of establishing up to 10$^8$ unique chemical signatures may
appear to be a hopeless proposition with current technology. But it
is worth noting that more than 60 of the chemical elements ($Z > 30$)
arise from n-capture processes. Let us suppose that half of these are
detectable for a given star. We would only need to be able to 
measure two distinct abundances for each of these elements in 
order to achieve 10$^9$ independent cells in ${\cal C}$-space.
If many of the element abundances are found to be {\it rigidly}
coupled, of course the parameter space would be much smaller.

It may not be necessary to measure as many as 30 elements if some
can be found which are highly decoupled and exhibit large relative
dispersions from star to star. Burris\etal\ (2000) demonstrate one 
such element pair, i.e. [Ba/Fe] and [Sr/Fe]. It is difficult at 
this stage to suggest which elements are most suited to chemical
tagging. In part, this depends on the precise details and mechanism
of formation of the n-capture elements at low [Fe/H]. 

The element abundances [X$_i$/Fe] show three main peaks at Z $\approx$
26, Z $\approx$ 52, and Z $\approx$ 78; the last two peaks are evident
in Fig.~\ref{fig3}. There have been
suggestions that the r-process gives rise to random abundance patterns
(\eg\ Goriely \& Arnould 1996) although this is not supported by new
observations of a few metal poor stars.  Heavy r-process elements
around the second peak compared to the Sun appear to show a universal
abundance pattern (Sneden\etal\ 2000; Cayrel\etal\ 2001;
Hill\etal\ 2002). However, Hill\etal\ find that the third peak and
actinide elements (Z $\geq$90) are decoupled from elements in the second
peak. We suspect that there may be a substantial number of suitable
elements ($\sim$10) which could define a sufficiently large parameter
space. 

Our ability to detect structure in ${\cal C}$-space depends on how
precisely we can measure abundance differences between stars. It may be
possible to construct a large database of differential abundances from
echelle spectra, with a precision of 0.05 dex or better; differential
abundances are preferred here to reduce the effects of systematic
error.

\section{Chemical trajectories}

Our simple picture assumes that a cloud forms with a unique chemical
signature, or that shortly after the cloud collapses, one or two
massive SN II enrich the cloud with unique yields which add to the
existing chemical signature. The low-mass population forms with this
unique chemical signature. If the star-formation efficiency is high
($\gta 30$\%), the star group stays bound although the remaining gas is
blown away. If the star-formation efficiency is low ($\lta 10$\%), the
star cluster disperses along with the gas. In a closed box model, the
dispersed gas reforms a cloud at a later stage.

In the closed box model, each successive generation of supernovae
produce stellar populations with progressive enrichments. These will
lie along a trajectory in ${\cal C}$-space which can be identified in
principle using minimum spanning tree methods (Sedgewick 1992).  The
overall distribution of the trajectories will be affected by
fundamental processes like the star formation efficiency, the star
formation timescale, the mixing efficiency, the mixing timescale, and
the satellite galaxy infall rate.

As we approach solar levels of metallicity in [Fe/H], the vast number 
of trajectories will converge. By [Fe/H] $\approx$ -2.5, AGB stars will
have substantially raised the s-process element abundances; by [Fe/H]
$\approx$ -1, Type Ia supernovae will have raised the Fe-group
abundances.  Star clusters that appear to originate at the same
location in this ${\cal C}$-space may simply reflect a common formation
site, \ie\ the resolution limit we can expect to achieve in
configuration space. The ability to identify common formation sites
rests on accurate differential abundance analyses
(Edvardsson\etal\ 1993; Prochaska\etal\ 2000).

Even with a well established family tree based on chemical trajectories
in the chemical ${\cal C}$-space, this information may not give a clear
indication of the original location within the protocloud or Galactic
component. This will come in the future from realistic baryon
dissipation models. Forward evolution of any proposed model must be able
to produce the chemical tree.

However, the ${\cal C}$-space will provide a vast amount of
information on chemical evolution history. It should be possible to
detect the evolution of the cluster mass function with cosmic time
(Kroupa 2002), the epoch of a starburst phase and/or associated mass
ejection of metals to the halo (Renzini 2001), and/or satellite infall
(Noguchi 1998).

As we go back in time to the formation of the disk, we approach the
chemical state laid down by population~III stars.  The lack of stars
below [Fe/H] $\approx$ -5 suggests that the protocloud was initially
enriched by the first generation of stars (Argast\etal\ 2000). However,
the apparent absence of any remnants of population~III remains a
puzzle: its stars may have had a top-heavy initial mass function, or
have dispersed into the intra-group medium of the Local Group.  If one
could unravel the abundances of heavy elements at the time of disk
formation, this would greatly improve the precision of
nucleo-cosmochronology.

\section{Candidates for chemical tagging}

Chemical tagging will not be possible for all stars.  In hot stars, 
our ability to measure abundances is reduced by the stellar rotation
and lack of transitions for many ions in the optical.
The ideal candidates are the evolved FGK stars that are numerous
(10\%) and intrinsically bright. These can be observed at echelle
resolutions ($R > 30,000$) over the full Gaiasphere. Moreover, giants
have deep, low density atmospheres that produce strong low-ionization
absorption lines compared to higher gravity atmospheres.  Even in the
presence of significant line blending, with sufficient signal, it
should be possible to derive abundance information by comparing the
fine structure information with accurate stellar synthesis models.
Detailed abundances of large numbers of F and G subgiants would be
particularly useful, if it becomes possible to make such studies,
because direct relative ages can be derived for these stars from their
observed luminosities.

It is not clear at what [Fe/H] the r-process elements become swamped by
the ubiquitous Fe-group and s-process elements.  At a resolution of
$R\sim 10^5$, many r-process elements can be seen in the solar
spectrum, although the signal-to-noise ratio of about 1000 is needed,
and even then the spectral lines are often badly blended (Kurucz 1991;
1995).  Travaglio\etal\ (1999) suggest that the s-process does not
become significant until [Fe/H] $\approx$ -1 because of the need for
pre-existing seed nuclei (Spite \& Spite 1978; Truran 1981), although
Pagel \& Tautvaisiene (1997) argue for some s-process production at
[Fe/H] $\sim$ -2.5.  Prochaska\etal\ (2000) detected Ba, Y and Eu in a
snapshot survey of thick disk G dwarfs in the solar neighbourhood with
-1.1 $\lta$ [Fe/H] $\lta$ -0.5. This survey only managed to detect a
few transitions in each element although their spectral coverage was
redward of 440nm with SNR $\approx$ 100 per pixel at $R\simeq 50,000$.
Longer exposures with $R\sim 10^5$ and spectral coverage down to 300nm
would have detected more heavy elements.

\section{Future progress}

In our view, observations of nucleosynthetic signatures of metal-poor
stars provide a cornerstone of near-field cosmology. Success in this
arena requires major progress across a wide front, including better
atomic parameters (Truran\etal\ 2001), improved supernova models,
better stellar synthesis codes and more realistic galaxy formation
models.  There are no stellar evolutionary models that lead to a
self-consistent detonation and deflagration in a core-collapse
supernova event or, for that matter, detonation in a thermonuclear
explosive event.  Realistic chemical production at the onset of the
supernova stage requires a proper accounting of a large number of
isotope networks (400$-$2500) that cannot be adequately simulated
yet.  Modern computers have only recently conquered relatively
simple $\alpha$ networks involving 13 isotopes. The inexorable march of
computer power will greatly assist here.  

There is also a key experimental front both in terms of laboratory
simulations of nucleosynthesis, and the need for major developments in
astronomical instrumentation. Many authors (\eg\ Sneden\etal\ 2001$b$)
have stressed the importance of greatly improving the accuracy of
transition probabilities and reaction rates for both heavy and light
ion interactions. This will be possible with the new generation
of high-intensity accelerators and radioactive-beam instruments
(K\"{a}ppeler\etal\ 1998; \qv Manuel 2000).

Progress on all fronts will require iteration between the different
strands. Already, relative r-process and $\alpha$ element abundances
for metal-poor stars have begun to constrain the yields for different
stellar masses and associated mass cuts of progenitor supernovae
(Mathews\etal\ 1992; Travaglio\etal\ 1998; Ishimaru \& Wanajo
2000).

Detailed high resolution abundance studies of large samples of galactic
stars will be crucial for the future of fossil astronomy.  But we stress
that in order to access a representative sample of the Gaiasphere,
this will require a new generation of ground-based instruments,
in particular, a multi-object echelle spectrograph with good blue
response on a large aperture telescope.  There is a real need for a
high-resolution spectrograph which can reach hundreds or even thousands
stars in a square degree or more. The Gemini Wide Field proposal currently
under discussion provides an opportunity for this kind of instrument
(S. Barden, personal communication).  The instrument will be expensive
and technically challenging, but we believe this must be tackled if we
are to ever unravel the formation of the Galaxy.

Could some of the residual inhomogeneities from prehistory have
escaped the dissipative process at an early stage? We may not know the
answer to this question with absolute certainty for many years. 
But it is an intriguing thought that one day we may be able to identify
hundreds or even thousands of stars throughout the Gaiasphere that were 
born within the same cloud as the Sun.

\bigskip
\noindent{\bf Acknowledgment.} 
I am deeply indebted to Ken Freeman for his inspiration and dialogue over
many years. This contribution is in honour of him, and in fact arose
out of our work on the 2002 Annual Reviews article.

\end{document}